# Graphyne: Hexagonal network of carbon with versatile Dirac cones

Bog G. Kim[*]
*Department of Physics, Pusan National University, Busan 609-735, Korea*

Hyoung Joon Choi[†]
*Department of Physics and IPAP, Yonsei University, Seoul 120-749, Korea*

We study $α$, $β$, and $γ$ graphyne, a class of graphene allotropes with carbon triple bonds, using a first-principles density-functional method and tight-binding calculation. We find that graphyne has versatile Dirac cones and it is due to remarkable roles of the carbon triple bonds in electronic and atomic structures. The carbon triple bonds modulate effective hopping matrix elements and reverse their signs, resulting in Dirac cones with reversed chirality in $α$ graphyne, momentum shift of the Dirac point in $β$ graphyne, and switch of the energy gap in $γ$ graphyne. Furthermore, the triple bonds provide chemisorption sites of adatoms which can break sublattice symmetry while preserving planar $sp^2$-bonding networks. These features of graphyne open new possibilities for electronic applications of carbon-based two-dimensional materials and derived nanostructures.

Recently there has been enormous interest in two-dimensional (2D) structures of carbon allotropes.[1-4] Among them, graphene has been a principal focus of research activity. Graphene is a 2D hexagonal network of carbon atoms, having strong triangular σ bonds of the $sp^2$ hybridized orbitals.[1-3] The electronic structure of graphene is characterized by the existence of the Dirac cones, where electron and hole spectra meet linearly at single points in the momentum space, called the Dirac points, and the charge carrier has the pseudospin parallel or antiparallel to its momentum.[3]

One of the biggest challenges in graphene is to manipulate its electronic conduction by opening an energy gap at the Dirac point[3,5-7] or by exploiting the pseudospin. Several approaches have been studied including sublattice symmetry breaking through an interaction with a substrate,[5,7] patterning into nanoribbons and nanomeshes,[6] and chiral symmetry breaking mechanism by 2D extension of Peierls distortion.[8] External stress was also considered, but recent calculations showed that the Dirac cones are very stable against external stress.[9-12]

Since the Dirac cone and the pseudospin originate from structural symmetry of graphene, other 2D carbon allotropes of similar structure may share the same property. An intriguing candidate is graphyne whose atomic structure was suggested by Baughman *et al.* in 1987.[13] They suggested various types of graphyne by inserting carbon triple bonds (-C≡C-) into C-C bonds in graphene, including three highly symmetric forms: $α$, $β$, and $γ$ graphyne.[13] Graphyne is named after yne carbon which denotes the carbon triple bond. While early *ab initio* calculations were focused on 2D and three-dimensional forms of graphyne without any report or discussion of Dirac cones,[14,15] very recent experimental success of graphdiyne synthesis[16] also motivated electronic structure calculations of the 2D graphdiyne and graphdiyne nanoribbons with strong interest.[17,18]

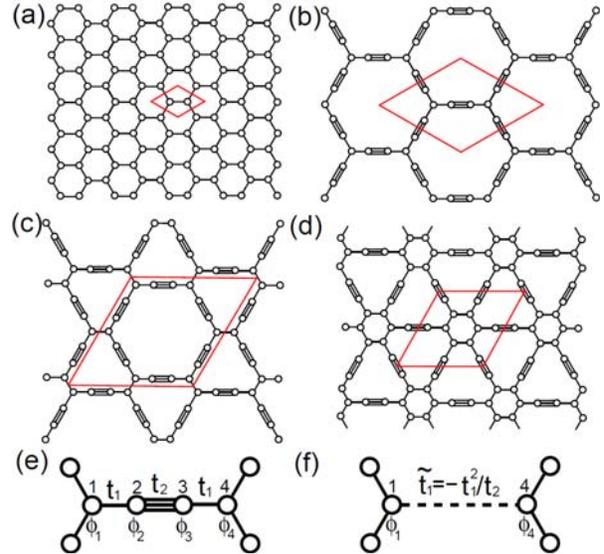

FIG. 1. (Color online) Schematic representations of (a) graphene, and (b) $α$, (c) $β$, and (d) $γ$ graphyne. Red quadrangles indicate unit cells. (e) The hopping matrix elements along a carbon triple bond in graphyne. The carbon atoms 1 and 4 are at vertices of a hexagon in graphyne. (f) Effective direct hopping matrix element between the two carbon atoms 1 and 4 in (e).

In this paper, we study the electronic structures of $α$, $β$, and $γ$ graphyne using first-principles density-functional calculations, reporting two and six Dirac cones in Brillouin zone (BZ) of $α$ and $β$ graphyne, respectively. In $α$ graphyne, the carbon triple bonds produce opposite chirality to that in graphene and they can also host adatoms that open an energy gap at the Dirac point by breaking the sublattice symmetry. While $γ$ graphyne is semiconducting at the optimized atomic structure, it also has Dirac cones when the carbon triple bonds are slightly elongated. We provide a simple tight-binding Hamiltonian which captures essential features of Dirac cones in graphyne in a unified way.

TABLE I. Optimized lattice constant (*a*) and lengths of carbon bonds for *α*, *β*, and *γ* graphyne. $d_t$ denotes the length of the triple bond and $d_{s1}$ and $d_{s2}$ denote lengths of two different single bonds. Note that there is only one kind of single bond for *α* graphyne.

| Structure | $a$ (Å) | $d_{s1}$ (Å) | $d_{s2}$ (Å) | $d_t$ (Å) |
|---|---|---|---|---|
| *α* graphyne | 6.9812 | 1.3995 | - | 1.2317 |
| *β* graphyne | 9.5004 | 1.3922 | 1.4633 | 1.2343 |
| *γ* graphyne | 6.8826 | 1.4070 | 1.4237 | 1.2214 |

We performed first-principles calculations with the generalized gradient approximation[19] to the density functional theory and the projector-augmented-wave method as implemented in VASP.[20,21] We regarded $2s^2 2p^2$ electrons in a carbon atom as valence electrons. Electronic wave functions are expanded with plane waves up to a kinetic-energy cutoff of 400 eV. Momentum-space integration is performed using a 7 × 7 × 1 Monkhorst-Pack *k*-point mesh centered at Γ for hexagonal structure. With the hexagonal symmetry imposed, lattice constants and internal coordinates were fully optimized until the residual Hellmann-Feynman forces became smaller than $10^{-3}$ eV/Å.

As shown in Fig. 1, graphene can be modified by inserting triple bonds of carbon (-C≡C-) while keeping its 2D hexagonal symmetry. Figure 1 shows the structures of (a) graphene and its yne modifications, (b) *α*, (c) *β*, and (d) *γ* graphyne.[13] The optimized structural parameters for graphyne are summarized in Table I. The most symmetric modification of graphene is *α* graphyne [Fig. 1(b)], where the carbon triple bond (-C≡C-) is inserted into every carbon bond in graphene. In the original proposal,[13] hexagons in *α* graphyne are not equilateral and it has resonance structures. In our calculation the optimized structure of *α* graphyne has equilateral hexagons only, and it is stable without any phonon mode of imaginary frequency. Figure 1(c) shows the structure of *β* graphyne, where carbon triple bonds are inserted into two thirds of C-C bonds in graphene, forming enlarged hexagons bridged by slightly elongated C-C bonds. In *γ* graphyne [Fig. 1(d)], the carbon triple bonds are inserted into only one third of C-C bonds in graphene, resulting in benzene rings connected by carbon triple bonds. The structural symmetry of *β* and *γ* graphyne is equivalent to that of graphene under the Kekule distortion.

All of graphyne are topologically equivalent to graphene if we count carbons at hexagonal vertices only. In graphyne, electrons hop from one vertex carbon to another either directly or through a carbon triple bond. The hopping through a carbon triple bond can be regarded as a renormalized direct hopping, as follows. Let us consider a simplified Hamiltonian at four carbon sites as shown in Fig. 1(e), where $\varphi_i$ is the *i*th-site wave function and $t_1$ and $t_2$ are the hopping matrix elements for the single and the triple bond, respectively. Hamiltonian equations at the carbon triple bond (that is, at sites 2 and 3) are

$$E\varphi_2 = V\varphi_2 + t_1\varphi_1 + t_2\varphi_3, \quad E\varphi_3 = V\varphi_3 + t_2\varphi_2 + t_1\varphi_4, \quad (1)$$

where *E* is the electron energy and *V* is the on-site energy. These equations yield

$$\varphi_2 = \frac{t_1(E-V)\varphi_1 + t_1 t_2 \varphi_4}{(E-V)^2 - t_2^2}, \quad \varphi_3 = \frac{t_1 t_2 \varphi_1 + t_1(E-V)\varphi_4}{(E-V)^2 - t_2^2}, \quad (2)$$

which are approximately $\varphi_2 \cong -(t_1/t_2)\varphi_4$ and $\varphi_3 \cong -(t_1/t_2)\varphi_1$ near the Fermi energy where $E \approx V$. Thus the hopping term from site 2 to 1 is $t_1\varphi_2 \cong -(t_1^2/t_2)\varphi_4 \equiv \tilde{t}_1\varphi_4$ and that from site 3 to 4 is $t_1\varphi_3 \cong -(t_1^2/t_2)\varphi_1 \equiv \tilde{t}_1\varphi_1$, so they can be regarded as a direct hopping between sites 1 and 4 with a renormalized matrix element

$$\tilde{t}_1 \equiv -(t_1^2/t_2), \quad (3)$$

as shown in Fig. 1(f). Remarkably, the renormalized hopping matrix element $\tilde{t}_1$ has the opposite sign to $t_1$, and its magnitude depends on detailed characteristics of the carbon triple bond. Using the renormalized hopping matrix element, graphyne reduces to graphene with a variation in the hopping matrix element.

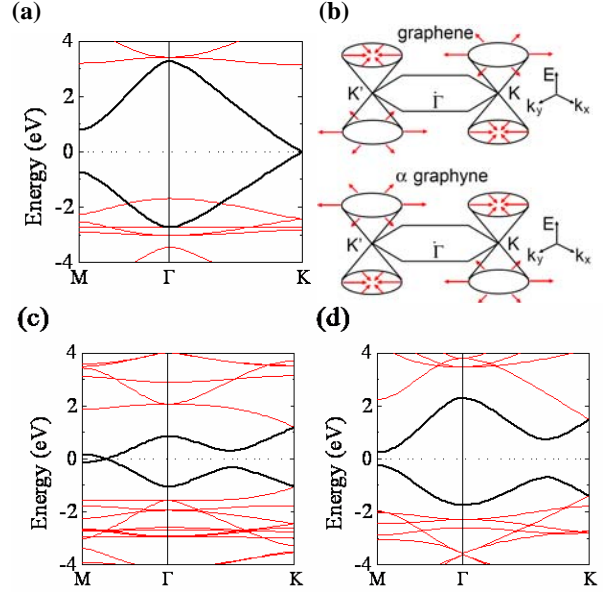

FIG. 2. (Color online) (a) Electronic band structure of *α* graphyne along *M*-Γ-*K* lines. (b) Comparison of Dirac cones and pseudospins in graphene and *α* graphyne. Red arrows indicate pseudospins. (c) Electronic band structure of *β* graphyne. (d) Electronic band structure of *γ* graphyne. In (a) and (c) Fermi energy is set zero. In (d) the middle of the energy gap is set zero.

Figure 2 shows electronic structures of $\alpha$, $\beta$, and $\gamma$ graphyne obtained by first-principles calculations. Except for a few nonpropagating flat bands from carbon triple bonds, $\alpha$ graphyne has almost the same electronic structure as graphene near $E_F$ [Fig. 2(a)], having Dirac cones at the $K$ and $K'$ points. In $\alpha$ graphyne, the Fermi velocity is $6.86\times10^5$ m/s [Fig. 2(a)], which is about 80 % of the Fermi velocity in graphene. Since the renormalized hopping matrix element $\tilde{t}_1$ has the opposite sign to $t_1$, the pseudospin in $\alpha$ graphyne has the opposite direction to that in graphene, as depicted in Fig. 2(b).

Dirac cones are also present in $\beta$ graphyne [Fig. 2(c)], but its electronic structure is quite different from that of graphene or $\alpha$ graphyne. Dirac cones in $\beta$ graphyne are not at the $K$ or $K'$ point, but they are located along $\Gamma$-$M$ lines, resulting in six Dirac cones in BZ. At each Dirac cone, the Fermi velocity is $4.45\times10^5$ m/s along the $\Gamma$-$M$ direction and $5.24\times10^5$ m/s perpendicular to the $\Gamma$-$M$ direction.

In the case of $\gamma$ graphyne, the energy gap is fully opened in the entire BZ [Fig. 2(d)] with a direct energy gap ($E_g$ = 0.471 eV) at the $M$ point. We find that this gap opening is highly dependent on the triple bond length ($d_t$) [Fig. 3(a)]. In equilibrium geometry, $d_t$ is 1.2214 Å and the lattice constant is 6.8826 Å. We checked the size of the energy gap of $\gamma$ graphyne while changing $d_t$ without changing the lattice constant [Fig. 3(b)]. As $d_t$ increases, the energy gap decreases linearly until $d_t$ = 1.2780 Å. For $d_t$ > 1.2780 Å, the energy gap eventually closes and Dirac cones appear in $\gamma$ graphyne [Fig. 3(c)]. The location of the Dirac cone is along the $M$-$\Gamma$ line and can be tuned by $d_t$. Figures 3(d) and 3(e) show the electronic wave functions at the valence band maximum which are plotted in a plane containing a carbon triple bond for two different $d_t$, respectively. With the equilibrium $d_t$ of 1.2214 Å, the wave function is localized at the carbon triple bond [Fig. 3(d)], whereas with the elongated $d_t$ of 1.2780 Å, the wave function is delocalized along the four carbon atoms [Fig. 3(e)]. These features show that the energy-gap opening in $\gamma$ graphyne is due to the Peierls instability occuring in a 2D fashion, which corresponds to the Kekule distortion.[8,22]

Having all the band structure information, we now discuss a unified picture of graphyne. As we mentioned above, all of graphyne are topologically equivalent to graphene if we count vertex carbon atoms only, and the electron hopping from a vertex site to another through a carbon triple bond can be regarded effectively as a direct hopping with a renormalized hopping matrix element. With this picture, the delocalized electronic states near the Fermi energy or the energy gap in graphyne can be described with a tight-binding Hamiltonian in a honeycomb lattice with renormalized nearest-neighbor hopping matrix elements ($t_{ij}$),[23-26]

$$H = \sum_{<ij>} t_{ij}\left(a_j^+ b_i + H.c.\right), \quad (4)$$

where $<ij>$ denotes summing over the nearest neighbors, and $a_j^+$ and $b_j^+$ are the operators creating an electron on the A and B sublattices, respectively [Fig. 4(a)]. Note that, in graphene, $t_{ij}$ is same for all three nearest neighbors ($t_0$).[3,24] With two vertex sites in a unit cell, $t_{ij}$ can have three different values $\tilde{t}_1$, $\tilde{t}_2$, and $\tilde{t}_3$, in general, depending on the directions of the nearest neighbors, $\vec{d}_1$, $\vec{d}_2$, and $\vec{d}_3$, respectively, as shown in Fig. 4(a). Using plane-wave-like linear combinations of $a_j^+$ and $b_j^+$ with a wave vector $\vec{k}$, the Hamiltonian can be expressed with a 2×2 matrix,

$$H(\vec{k}) = \begin{pmatrix} 0 & f(\vec{k}) \\ f(\vec{k})^* & 0 \end{pmatrix}, \quad (5)$$

with $f(\vec{k}) = \tilde{t}_1 \exp(-i\vec{k}\cdot\vec{d}_1) + \tilde{t}_2 \exp(-i\vec{k}\cdot\vec{d}_2) + \tilde{t}_3 \exp(-i\vec{k}\cdot\vec{d}_3)$. This yields two energy bands $E(\vec{k}) = \pm|f(\vec{k})|$. It is straightforward to show that if the three values $|\tilde{t}_1|$, $|\tilde{t}_2|$, and $|\tilde{t}_3|$ can form a triangle, $f(\vec{k})$ is zero at a certain $k$,[25,26] where the upper and the lower energy band touch each other with a linear dispersion relationship, forming a Dirac cone. In the other case that $f(\vec{k})$ is not zero at any given $k$, the energy gap should be open in the entire BZ.

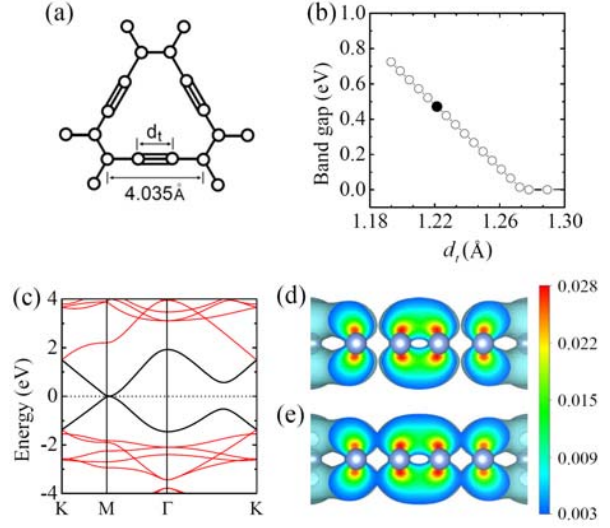

FIG. 3. (Color online) (a) Definition of the triple bond length ($d_t$) of $\gamma$ graphyne. (b) Energy gap in $\gamma$ graphyne as a function of $d_t$. (c) Electronic band structure of $\gamma$ graphyne for $d_t$ = 1.2780 Å. (d) Electronic wave function at the valence band maximum of $\gamma$ graphyne for the equilibrium $d_t$ of 1.2214 Å. (e) Electronic wave function at the valence band maximum of $\gamma$ graphyne for the elongated length $d_t$ = 1.2780 Å.

In the case of $\alpha$ graphyne, the three hopping matrix elements are equal to one another ($\tilde{t}_1 = \tilde{t}_2 = \tilde{t}_3$), so $f(\vec{k})$ is zero at the $K$ and $K'$ points [Fig. 4(b)] regardless of the value of $\tilde{t}_1$, producing Dirac cones at the $K$ and $K'$ points [Fig. 2(a)]. In the case of $\beta$ and $\gamma$ graphyne, only two hopping matrix elements are equal to each other (say $\tilde{t}_1 = \tilde{t}_2 \neq \tilde{t}_3$). So, if $|\tilde{t}_3|$ is less than $2|\tilde{t}_1|$, the Dirac cone exists away from the $K$ or $K'$ point, as in $\beta$ graphyne [Fig 2(c)]. On the other hand, if $|\tilde{t}_3|$ is greater than $2|\tilde{t}_1|$, the Dirac point disappears [Figs. 4(c) and 4(d)], as in $\gamma$ graphyne [Fig. 2(d)]. Although this tight-binding model has only two vertex carbon sites in a unitcell, it captures essential features of $\beta$ and $\gamma$ graphyne that has six vertex carbon sites in each unitcell. We note that, in $\gamma$ graphyne, the triple bond length is a tuning parameter for the hopping matrix element $|\tilde{t}_3|$, switching on-and-off the energy gap at the Dirac point, as shown in Figs. 2(d) and 3(c).

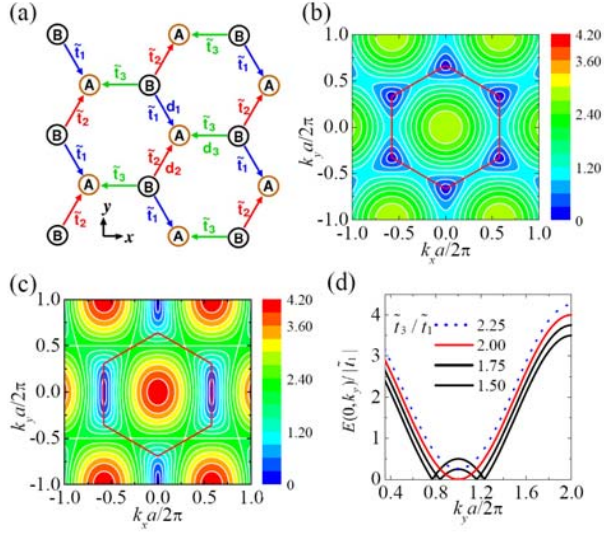

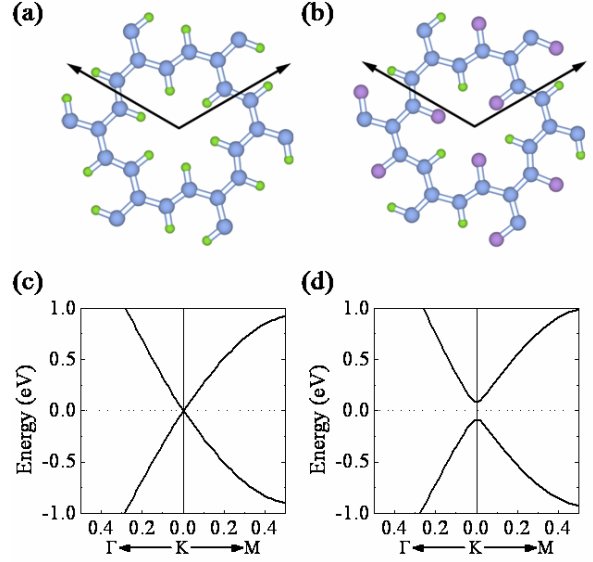

FIG. 4. (Color online) (a) Honeycomb lattice geometry with three nearest-neighbor hopping matrix elements ($\tilde{t}_1$, $\tilde{t}_2$, $\tilde{t}_3$) and three nearest-neighbor relative positions ($\vec{d}_1$, $\vec{d}_2$, $\vec{d}_3$). A and B denote two sublattices. (b) 2D plot of the energy dispersion, $E(k_x, k_y)/|\tilde{t}_1|$ with three equivalent hoppings ($\tilde{t}_1 = \tilde{t}_2 = \tilde{t}_3$). (c) 2D plot of the energy dispersion $E(k_x, k_y)/|\tilde{t}_1|$ with $\tilde{t}_1 = \tilde{t}_2$ and $\tilde{t}_3 = 2.1 \times \tilde{t}_1$. (d) The dispersion $E(0, k_y)/|\tilde{t}_1|$ for various values of $\tilde{t}_3/\tilde{t}_1$. In (d) the energy-gap opening can be clearly seen when $\tilde{t}_3/\tilde{t}_1 > 2$. In (b)-(d) the lattice constant $a$ is $\sqrt{3}$ times the C-C bond length.

Another important feature of graphyne is its potential for energy-gap control at the Dirac point by sublattice symmetry breaking. While maintaining planar $sp^2$-bonding structure of carbon atoms, one can easily modify electronic structure of graphyne by adding atoms to the carbon triple bond. Figures 5(a) and 5(b) show optimized geometry of hydrogenated $\alpha$ graphyne and hydrogen-fluorine-added $\alpha$ graphyne, respectively. Introduction of two hydrogen atoms added to each triple bond deforms the linear carbon chain [Fig. 5(a)], but it does not change the electronic structure significantly because sublattice symmetry is preserved and renormalization of the hopping matrix element does not affect the Dirac cone [Fig. 5(c)]. On the contrary, when one replaces half of hydrogen atoms with fluorine atoms, the adatoms break the sublattice symmetry [Fig. 5(b)], which is equivalent to introducing two different diagonal terms ($E_A$ and $E_B$ instead of zero) into the Hamiltonian of Eq. (5). This broken symmetry results in an energy gap larger than $|E_A - E_B|$ in any $k$ point, opening an energy gap at the Dirac point [Fig. 5(d)].

FIG. 5. (Color online) Optimized atomic structures of (a) $H_2$ added $\alpha$ graphyne and (b) HF added $\alpha$ graphyne. Blue, green, and purple dots represent C, H, and F, respectively. (c) Electronic band structure of (a). (d) Electronic band structure of (b). Note that the energy gap opens in (d).

In conclusion, we have studied the optimized structures and electronic structures of $\alpha$, $\beta$, and $\gamma$ graphyne with first-principles density-functional calculations and a tight-binding method. Dirac cones are found present in $\alpha$ and $\beta$ graphyne. In $\alpha$ graphyne, pseudospin is in the opposite direction to that in graphene. While $\gamma$ graphyne is semiconducting due to Kekule-distortion effect, Dirac cones appear if the carbon triple bond is elongated. A tight-binding Hamiltonian for a honeycomb lattice with three different hopping matrix elements can explain all the important physics of graphyne. Furthermore, the carbon triple bond in graphyne can be easily modified by attaching hydrogen or halogen atoms while maintaining 2D planar hexagonal symmetry and this modification can tune parameter for the energy gap at the Dirac point. Our present work demonstrates that 2D graphyne has rich physics in its electronic structures, which would be extended further in related nanostructures such as graphyne nanotubes, nanoribbons, quantum dots, and junctions.

We thank R. Baughman for helpful discussions and encouragement in initial state of this work. This work was supported by NSF of Korea (KRF-2011-0006256, KRF-2011-0002511, and 2011-0018306). Computational resources have been provided by KISTI Supercomputing Center (Project No. KSC-2011-C2-04 and No. KSC-2011-C1-20).